\documentstyle[aps,pra,eqsecnum,preprint,tighten]{revtex}
\input{epsf}

\def\bold#1{\mbox{\boldmath $#1$}}
\def\beq{\begin{equation}}
\def\eeq{\end{equation}}
\def\beqa{\begin{eqnarray}}
\def\eeqa{\end{eqnarray}}
\def\rg{\rangle}
\def\lg{\langle}
\def\rar{\rightarrow}
\def\nn{\nonumber}
\def\ot{\otimes }
\def\Ph{{\rm Ph}}
\def\Tr{{\rm Tr}}

\def\ad{^\dagger }
\def\al{\alpha }
\def\bt{\beta }
\def\dl{\delta }

\def\gm{\gamma }
\def\kp{\kappa}
\def\half{{\textstyle {1\over2}}}

\def\lam{\lambda }

\def\dbd{\bold{d}}

\def\Mbd{\bold{M}}

\def\rbd{\bold{r}}
\def\si{\sigma}

\def\zt{\zeta }

\def\Ac{{\cal A}}
\def\Bc{{\cal B}}
\def\Fc{{\cal F}}
\def\Gc{{\cal G}}
\def\Hc{{\cal H}}

\begin{document}

\title{Two Qubit Copying Machine for Economical Quantum Eavesdropping}

\author{Chi-Sheng Niu\thanks{Electronic mail: cn28+@andrew.cmu.edu} and
Robert B. Griffiths\thanks{Electronic mail: rgrif@cmu.edu}\\ Department of
Physics, Carnegie Mellon University,\\ Pittsburgh, PA 15213}

\date{Version of 4/29/99}

\maketitle

\begin{abstract}
	We study the mapping which occurs when a single qubit in an arbitrary
state interacts with another qubit in a given, fixed state resulting in some
unitary transformation on the two qubit system which, in effect, makes two
copies of the first qubit.  The general problem of the quality of the resulting
copies is discussed using a special representation, a generalization of the
usual Schmidt decomposition, of an arbitrary two-dimensional subspace of a
tensor product of two 2-dimensional Hilbert spaces.  We exhibit quantum
circuits which can reproduce the results of any two qubit copying machine of
this type.  A simple stochastic generalization (using a ``classical'' random
signal) of the copying machine is also considered.  These copying machines
provide simple embodiments of previously proposed optimal eavesdropping schemes
for the BB84 and B92 quantum cryptography protocols.
\end{abstract}

		\section{introduction}\label{s1}

	Recent advances in quantum computation, and proposals for quantum
cryptographic schemes, have led to a renewed interest in quantum information
theory: how information is stored, processed, corrupted, and preserved from
corruption in situations where quantum effects play an essential role.  While
classical information theory provides a starting point for a quantum theory of
information, it is clear that classical ideas are not sufficient, but what
should replace them is at present far less clear.

	As in any area of theoretical physics, two very different approaches
are possible, as well as a broad continuum in between: one can search for very
general results applicable to any quantum system, or one can work out specific
models and simple examples.  The present paper belongs to the second category:
we are interested in what happens when two qubits, the simplest conceivable
carriers of quantum information, interact with each other.  Qubits $a$ and $b$
---think of them as spins of two particles of spin one half---correspond to
two-dimensional Hilbert spaces $\Ac$ and $\Bc$.  We suppose that they interact
(``scatter'') during a finite time interval, which results in a unitary
transformation on the tensor product space $\Ac\ot\Bc$, after which they
separate from each other and can be subjected to various measurements.

	So far as we know, no one has worked out a complete characterization
(whatever that might be) of the things that can happen during such a unitary
transformation on two qubits.  The present paper represents only one step along
the way towards addressing this general problem.  We ask, and answer, the
following question: Suppose the $b$ qubit is initially in some arbitrary but
fixed state $|b\rg$, and the unitary transformation $U$ is is also arbitrary
but fixed.  How does what emerges from the interaction of the two qubits depend
upon the initial state of the $a$ qubit?  In particular, what might one learn
by carrying out measurements on the $a$ and $b$ qubits once their interaction
is over?

	A helpful way to view this question is in terms of quantum copying: the
two qubits which emerge can in some sense be thought of as ``copies'' of the
$a$ qubit made by a ``copying machine'' constituted by the fixed $b$ qubit
together with the fixed unitary transformation.  An essential aspect of the
quantum copying problem is that the quantum state to be copied is (initially)
unknown to the operator of the copying machine, who thus has to follow a fixed
protocol, corresponding in our case to the fixed $|b\rg$ and $U$.  It then
follows from the no-cloning theorem \cite{r1} that no copying machine can make
perfect copies of all incoming states, if they are not all orthogonal to each
other. 

	In a previous paper \cite{r2} we discussed in very general terms the
problem of producing two optimal copies of a single qubit, where optimality is
defined relative to a particular ``distinguishability'' measure of copy
quality.  The present article discusses the interaction of two qubits as a
two-qubit copying machine which, although less flexible than one which employs
a third ``ancillary'' qubit, still satisfies the optimality criteria of
\cite{r2}, and we show that some additional flexibility can be achieved using a
stochastic version of the two-qubit copier.  However, this two qubit copier
cannot serve as a ``universal'' copier in the sense defined in \cite{r2b}, for
reasons pointed out in Sec.~III of that paper.  (As the precise definition of
optimality is a bit lengthy, and is not needed to understand the present paper,
we refer the reader to \cite{r2} for the details, and for additional comments
on how our work is related to other literature on quantum copying.)

	Making an approximate copy for himself while leaving another reasonably
good copy behind is one way to state the problem faced by an eavesdropper
trying to extract information from a quantum channel without producing the sort
of noise which will alert the legitimate users of the channel to his presence.
Various optimal eavesdropping strategies have been proposed for two quantum
cryptographic protocols, BB84 \cite{r3} and B92 \cite{r4}.  In the case of
BB84, the previous proposal \cite{r5} required the use of three qubits (two in
addition to the one carrying the original signal); we show that a two qubit
copier suffices.  In the case of B92 it has been argued previously
\cite{r6,r6b}, though not conclusively proved, that two qubit eavesdropping can
produce optimal results.  Our contribution is to show that this can be done
using an even simpler, and presumably cheaper, quantum circuit.

	The interaction of two qubits also provides a simple example of a
``decoherence'' process in which quantum information, thought of as present
initially in the $a$ qubit, is (partially) lost through interaction with the
``environment'', in this case the $b$ qubit.  While such a simple environment
does {\it not} allow one to model the most general decoherence process possible
for a single qubit, it is nonetheless worth exploring precisely what it does
allow, a question which is answered in the present paper.

	The basic strategy which we employ for studying the problem,
Sec.~\ref{s2}, is to note that a unitary operator $U$ on the tensor product
$\Ac\ot\Bc$ maps the two-dimensional subspace $\Ac\ot | b\rg$, for a given,
fixed $|b\rg$, onto a two-dimensional subspace $\Gc$ of $\Ac\ot\Bc$.  This
mapping from $\Ac$ to $\Ac\ot\Bc$ is an isometry (it preserves the inner
product), and the task we face is, in essence, that of understanding and
classifying such isometries.  The classification begins by noting that the
subspace $\Gc$ can be characterized by two parameters, according to a theorem
stated in Sec.~\ref{s2}, which allows us to pick a basis for $\Gc$ in a
particularly convenient ``canonical'' form, generalizing the usual Schmidt
representation for a pure state.  With the help of this canonical basis we can
express any isometry as a combination of a ``canonical'' isometry of a
particularly simple structure, together with a series of one-qubit operations,
that is, unitary transformations on $\Ac$ and on $\Bc$.  The canonical isometry
can then be understood in geometrical terms using a Bloch sphere
representation, Sec.~\ref{s3}, which is also useful when considering the
additional possibilities represented by a {\it stochastic} copying machine,
Sec.~\ref{sx}.  Both the canonical isometry and any other isometry can be
easily implemented using simple quantum circuits, as shown in Sec.~\ref{s4}, to
produce copying machines, including stochastic copying machines if one allows
certain gates to be controlled by a stochastic ``classical'' signal.  In the
cryptographic context discussed in Sec.~\ref{s6}, these copying machines can
serve as simple eavesdropping devices.  A summary of our results, and an
indication of some open problems, is presented in Sec.~\ref{s7}

	\section{Isometries and Two-Dimensional Subspaces}
\label{s2}

	\subsection{Canonical basis for a two-dimensional subspace of
$\Ac\ot\Bc$}
\label{s2a}

	We are interested in the four-dimensional Hilbert space
\beq
  \Hc=\Ac\ot\Bc,
\label{e2.1}
\eeq
where $\Ac$ and $\Bc$ are the two-dimensional spaces associated with the qubits
$a$ and $b$.  Given an arbitrary, but fixed, initial state $|b\rg$ in $\Bc$ for
the $b$ qubit, and a fixed unitary interaction $U$ mapping $\Hc$ to itself, we
define the isometry $V:\Ac\rar \Hc$ by means of the equation
\beq
 V|\al\rg = U(|\al\rg\ot | b\rg).
\label{e2.2}
\eeq
Thus, $V$ tells us what happens to an arbitrary state $|\al\rg$ of the $a$
qubit if we assume that $|b\rg$ and $U$ are fixed.  The name ``isometry'' is
appropriate, for $V$ preserves inner products,
\beq
 (V|\al\rg)\ad (V|\al'\rg) = \lg\al | V\ad V |\al'\rg = \lg\al | \al'\rg.
\label{e2.3}
\eeq
It is evident that $V$ maps $\Ac$ onto some two-dimensional subspace $\Gc$ of
$\Hc$.

	Using the well-known Schmidt result, an arbitrary one-dimensional
subspace $\Fc$ of $\Ac\ot\Bc$ can be characterized in the following way.  Given
a vector $|f\rg$ of unit length in $\Fc$, there are orthonormal bases
$\{|a_0\rg,|a_1\rg\}$ and $\{|b_0\rg,|b_1\rg\}$ of $\Ac$ and $\Bc$ such that
\beq
 |f\rg = \mu |a_0 b_0\rg + \bar\mu |a_1 b_1\rg,
\label{e2.4}
\eeq
where $|a_0 b_0\rg$ means $|a_0\rg\ot| b_0\rg$, and the phases of the basis
states can always be chosen so that $\mu$ and $\bar\mu$ are both non-negative
real numbers.  Note that either $\mu$ or $\bar\mu$, the sum of whose squares is
one, can be thought of as giving an intrinsic ``geometrical'' characterization
of the subspace $\Fc$, that is, one that does not depend upon the choice of
bases in $\Ac$ and $\Bc$, or the phase of $|f\rg$. To be sure, relabeling the
bases allows the interchange of $\mu$ and $\bar\mu$, so one can always restrict
$\mu$ to lie between $0$ and $1/\sqrt{2}$.  It then serves as a sort of measure
of ``entanglement'' of the subspace $\Fc$.
	A three-dimensional subspace of $\Ac\ot\Bc$ can be characterized in the
same way by applying the above argument to its orthogonal complement, since the
latter is a one-dimensional space.  An analogous result for a two-dimensional
subspace is the following:

	{\bf Theorem.} Let $\Gc$ be a two-dimensional subspace of $\Ac\ot\Bc$,
where $\Ac$ and $\Bc$ are two-dimensional complex Hilbert spaces.  Then there
are orthonormal bases $\{|a_0\rg,|a_1\rg\}$, $\{|b_0\rg,|b_1\rg\}$, and
$\{|g_0\rg,|g_1\rg\}$ of $\Ac$, $\Bc$, and $\Gc$, respectively, such that
\beqa
 |g_0\rg &=& \mu |a_0 b_0\rg + \bar \mu |a_1 b_1\rg,
\nn\\
 |g_1\rg &=& \nu |a_0 b_1\rg + \bar \nu |a_1 b_0\rg,
\label{e2.5}
\eeqa
with (in general complex) coefficients satisfying
\beq
 |\mu|^2 +|\bar\mu|^2 = 1 = |\nu|^2 +|\bar\nu|^2.
\label{e2.6}
\eeq
Furthermore, the bases may always be chosen in such a way that
$\mu,\bar\mu,\nu,\bar\nu$ are real and non-negative.

	In what follows we shall refer to the basis $\{|g_0\rg,|g_1\rg\}$ as
the {\it canonical basis} of $\Gc$, and (\ref{e2.6}) as the {\it canonical
representation} of $\Gc$.  While there ought to be a simple, elegant proof, we
have not found one; App.~\ref{saa} contains our inelegant demonstration. It
makes use of the following result, which we shall want to refer to later:

	{\bf Lemma.} Any two-dimensional subspace of the tensor product
$\Ac\ot\Bc$ of two two-dimensional complex Hilbert spaces contains at least one
non-zero product vector of the form $|g\rg=|\al\rg\ot |\bt\rg$.

	In fact, the ``generic'' two-dimensional subspace $\Gc$ contains two
linearly-independent, but not mutually orthogonal, product vectors, which play
a role in our proof of the theorem, and have a certain geometrical significance
as will be explained below in Sec.~\ref{s3}.

	It is possible to represent the same subspace $\Gc$ using alternative
choices of the coefficients in (\ref{e2.5}), provided the corresponding bases
(of $\Ac$, $\Bc$, and $\Gc$) are appropriately modified.  The corresponding
symmetry operations on the coefficients are discussed in App.~\ref{sab}.  For
certain purposes it is convenient to choose a trigonometric representation in
terms of the two angles $\zt$ and $\eta$:
\beqa
 |g_0\rg &=& \cos(\zt/2) |a_0 b_0\rg + \sin(\zt/2) |a_1 b_1\rg,
\nn\\
 |g_1\rg &=& \cos(\eta/2) |a_0 b_1\rg + \sin(\eta/2) |a_1 b_0\rg.
\label{e2.7}
\eeqa
where the factors of $1/2$ are not essential, but convenient in terms of the
geometry of Bloch sphere representations, Sec.~\ref{s3}.  The analysis of
App.~\ref{sab} shows that one can always choose $\zt$ and $\eta$ to lie in the
region
\beq
 0 \leq \zt \leq \eta,\quad \zt+\eta \leq \pi.
\label{e2.8}
\eeq

	While the proof of the theorem given in App.~\ref{saa} is constructive,
it is not a particularly convenient way in practice to find the canonical
basis.  A simpler approach, which is satisfactory except for certain degenerate
cases, is described in App.~\ref{sac}.

	\subsection{Canonical form for $V$}
\label{s2b}

	The isometry $V$ in (\ref{e2.2}) maps all of $\Ac$ onto a
two-dimensional subspace $\Gc$ which, by the theorem, possesses a canonical
representation in the form (\ref{e2.5}).  We now use this canonical
representation to define a {\it canonical form} $V_c$ corresponding to the
isometry $V$.  Suppose that {\it standard} orthonormal bases $\{|0_a\rg
|1_a\rg\}$ and $\{|0_b\rg |1_b\rg\}$ are given for $\Ac$ and $\Bc$.  Think of
them as defined by some convenient convention; e.g., $|0\rg$ means that the
spin of the spin-half particle is in the $z$ direction. Relative to this basis
choice we define $V_c: \Ac\rar \Ac\ot\Bc$ through the equations
\beqa
 V_c |0\rg &=& \mu |0 0\rg + \bar \mu |1 1\rg = \cos(\zt/2) |0 0\rg +
\sin(\zt/2) |1 1\rg,
\nn\\
 V_c |1\rg &=& \nu |0 1\rg + \bar \nu |1 0\rg= 	\cos(\eta/2) |0 1\rg +
\sin(\eta/2) |1 0\rg,
\label{e2.9}
\eeqa
where the values of $\mu$, etc.~are the same as those in (\ref{e2.5}) and
(\ref{e2.7}), and $|0 1\rg$ is short for $|0_a 1_b\rg$: the $a$ qubit label is
to the left of the $b$ qubit label.

	The isometry $V$ is related to its canonical form through the equation
\beq
 V = (S_a\ot S_b) V_c S_o,
\label{e2.10}
\eeq
where $S_o$ and $S_a$ are one qubit unitary operations on $\Ac$, and $S_b$ is a
one qubit unitary operation on $\Bc$, defined in the following way.  Define the
vectors
\beq
 |a'_j\rg = V\ad |g_j\rg,
\label{e2.11}
\eeq
in $\Ac$; that is, $V$ maps $|a'_j\rg$ onto the canonical basis vector
$|g_j\rg$ of $\Gc$.  Because $V$ is an isometry, $\{|a'_0\rg, |a'_1\rg\}$ is an
orthonormal basis of $\Ac$.  If three one-qubit unitary operators are defined
by
\beq
 S_o |a'_j\rg = |j_a\rg,\quad S_a |j_a\rg = |a_j\rg,\quad S_b |j_b\rg =
|b_j\rg,
\label{e2.12}
\eeq
for $j=0$ and $1$, it is easily checked that (\ref{e2.10}) is satisfied.

	Just as the two parameters $\mu$ and $\nu$ can be thought of as
providing an intrinsic ``geometrical'' characterization of a two-dimensional
subspace of $\Ac\ot\Bc$, one which does not depend upon the the choice of
bases, similarly, $V_c$, which depends on the same two parameters, gives a sort
of intrinsic characterization of that part of the isometry $V$ that requires an
interaction between two qubits.  This is because one can think of $S_o$ and
$S_a$ in (\ref{e2.10}) as unitary operations applied to the $a$ qubit before
and after it interacts with the $b$ qubit, acting in effect as coordinate
transformations, and $S_b$ as a similar unitary operation or coordinate
transformation applied to the $b$ qubit after it has ceased interacting with
the $a$ qubit.  That there is no transformation on the $b$ qubit before the
interaction simply reflects the fact that the the initial state $|b\rg$
contributes to determining the canonical form $V_c$, as will be apparent when
we consider specific circuits in Sec.~\ref{s4}.

	A general isometry $V$ depends, if we ignore the overall phase, on 11
real parameters: 2 for $V_c$ and 3 for each of the one-qubit unitary
transformations in (\ref{e2.10}).  Hence it is very helpful to be able to
understand its essential features using just the two parameters which enter
$V_c$.  We shall make use of this simplicity in discussing the Bloch sphere
representation in Sec.~\ref{s3}, and constructing quantum circuits in
Sec.~\ref{s4}.
 
	\section{Bloch sphere representation}
\label{s3}

	The Bloch sphere representation provides a convenient way of thinking
about the isometries which interest us in an intuitive, geometrical way.  In
this language, the state of a single qubit is represented by a density matrix
$\rho$ in the form
\beq
 \rho = \half (\si_0 + \bold{r\cdot\si}),
\label{e3.1}
\eeq
where $\si_0$ is the identity operator and the $\si_j$ for $j >0$ are the usual
Pauli matrices in the standard basis (of the $a$ or $b$ qubit) with columns in
the order $|0\rg, |1\rg$.  Or, in terms of dyads,
\beqa
\si_1=\si_x&=&|0\rg\lg 1|+|1\rg\lg 0|,\nn\\
\si_2=\si_y&=&i|1\rg\lg 0|-i|0\rg\lg 1|,\nn\\
\si_3=\si_z&=&|0\rg\lg 0|-|1\rg\lg 1|.
\label{e3.2}
\eeqa
In (\ref{e3.1}), $\rbd=(r_1,r_2,r_3)$ is a real vector of length less than one,
for a mixed state, or equal to one, for a pure state.  For example,
$\rbd=(0,1,0)$ corresponds to a spin in the $y$ direction, $S_y=1/2$, in the
usual spin-half notation, whereas $\rbd=(0,0,0)$ is the completely mixed state
for which $S_y$, or any other spin component, takes the values $\pm 1/2$ at
random.

	If the initial state of the $a$ qubit is represented by $\rho^{in}$
corresponding to $\rbd^{in}$, the reduced density matrices of the $a$ and $b$
qubits after the interaction giving rise to the isometry $V$ are
\beq
\rho^a={\rm Tr}_\Bc \left[V\rho^{in}V^{\dag}\right],\quad
\rho^b={\rm Tr}_\Ac \left[V\rho^{in}V^{\dag}\right],
\label{e3.3}
\eeq
corresponding to $\rbd^a$ and $\rbd^b$.  The vectors $\rbd^a$ and $\rbd^b$ are
related to $\rbd^{in}$ through affine transformations,
\beq
\rbd^f=\Mbd^f\cdot\rbd^{in}+\dbd^f,
\label{e3.4}
\eeq
where $f$ stands for $a$ or $b$, $\Mbd^f$ is a $3\times 3$ real matrix, and
$\dbd^f$ a real three vector. (In equations of the form (\ref{e3.4}), the
vectors are column vectors, but we will generally write down the corresponding
row vectors, as in (\ref{e3.7}).)

	For an isometry in the canonical form $V_c$, one has the simple
expressions
\beq
\Mbd^a=\left(\matrix{\sin \gm &0            & 0 \cr
		0 	 &\sin \dl & 0 \cr 		0 & 0 & \sin \gm \sin
\dl }\right),
\label{e3.5}
\eeq
\beq
\Mbd^b=\left(\matrix{  \cos \dl  &0          & 0 \cr
		 0 &\cos \gm &0 \cr 		 0 &0 &\cos \gm \cos \dl
}\right),
\label{e3.6}
\eeq
and
\beqa
\dbd^a&=&(0,0,\cos \gm \cos \dl),\nn\\
\dbd^b&=&(0,0,\sin \gm \sin \dl ).\label{e3.7}
\eeqa
where $\gm$ and $\dl$ are defined by
\beq
 \gm=(\eta+\zt)/2,\quad \dl=(\eta-\zt)/2
\label{e3.8}
\eeq
in terms of the angles which appear in (\ref{e2.7}).

\begin{figure}
\epsfxsize=10truecm
\hspace*{3cm}\epsfbox{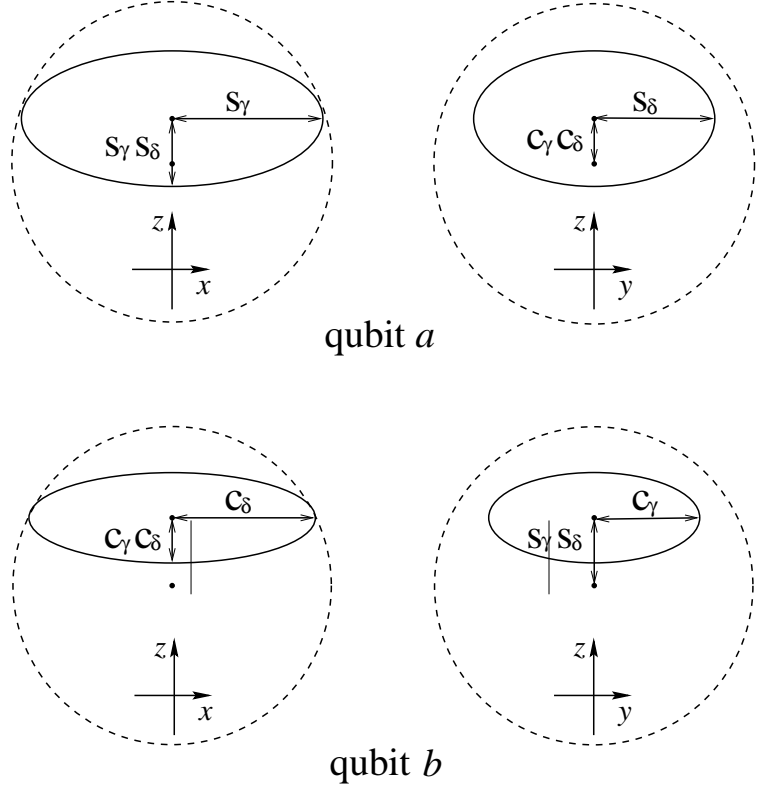}\vspace{.5cm}
\caption{
The ellipsoids corresponding to the two output qubits, viewed along the $x$ and
$y$ directions. Here $s_\gm$ $c_\gm$ $s_\dl$, and $c_\dl$ are all positive and
represent $\sin \gm$, $\cos \gm$, $\sin \dl$, and $\cos \dl$ respectively; the
figures are drawn assuming that $\sin \gm$ is larger than $\sin \dl$.  If some
of these quantities are negative, the ellipsoids may be displaced in the $-z$
direction.}
\label{f1}
\end{figure}

	The maps (\ref{e3.4}) provide a geometrical way of describing the state
evolution corresponding to the isometry $V$ as it produces ``copies'' of the
input state in the outgoing $a$ and $b$ qubits. In particular, the unit Bloch
sphere of possible pure states for the $a$ qubit before the interaction is
mapped into ellipsoids inside the Bloch spheres of the two qubits after the
interaction.  The principal semi-axes of these ellipsoids in three orthogonal
directions are given by the absolute values of the corresponding singular
values of $\Mbd^a$ and $\Mbd^b$, the diagonal elements of (\ref{e3.5}) and
(\ref{e3.6}) in the case of the canonical isometry $V_c$, and the vectors
$\dbd^a$ and $\dbd^b$ are the displacements of these ellipsoids from the
centers of the corresponding Bloch spheres.  The situation is illustrated for
$V_c$ in Fig.~\ref{f1}.

	As is evident from (\ref{e3.5}) and (\ref{e3.6}), for each ellipsoid
the length of the shortest semi-axis, parallel to $z$ in the case of $V_c$, is
the product of the lengths of the other two semi-axes, and is also the distance
the center of the {\it other} ellipsoid is displaced from the center of the
Bloch sphere.  Also, the square of the longest semi-axis for one ellipsoid plus
that of the second-longest axis for the other ellipsoid sum up to one.  These
constraints ensure that each ellipsoid touches the corresponding Bloch sphere
at two points, see Fig.~\ref{f1}, which represent pure states.  These points
correspond to the two linearly independent product states found in a
two-dimensional subspace $\Gc$ as discussed in App.~\ref{saa}.  In the special
case in which $\gm=\dl$ (modulo $\pi$), the two larger semi-axes are of equal
length, and the points where the ellipsoids touch the corresponding Bloch
spheres coalesce to a single point, apart from the special case in which one
ellipsoid is the entire Bloch sphere and the other is only a point.  

	A pair of ellipsoids whose semi-axes are given by the diagonal elements
in (\ref{e3.5}) and (\ref{e3.6}) satisfy the optimal copying conditions set
forth in \cite{r2}.  This can be shown by defining the quantities
\beqa
\bt_0 = \cos(\dl/2)\cos(\gm/2),\quad \bt_1 = \cos(\dl/2)\sin(\gm/2),
\nn\\
\bt_2 = \sin(\dl/2)\cos(\gm/2),\quad \bt_3 = \sin(\dl/2)\sin(\gm/2),
\label{e3.9}
\eeqa
and inserting them into (4.9) and (4.17) of \cite{r2} to obtain the elements on
the diagonal of (\ref{e3.6}) as a three-vector $\bold{b}$, and those on the
diagonal of (\ref{e3.5}) as a three vector $\bold{c}$.  Since, in addition, the
components of $\bold{b}$ satisfy (6.24) of \cite{r2}, $\bold{b}$ and $\bold{c}$
form an optimal pair in the notation of Sec.~VI A of \cite{r2}, which means
that the copying process corresponding to  the isometry $V_c$ is optimal.

	The general isometry $V$ is related to the corresponding canonical
isometry $V_c$ through (\ref{e2.10}).  In geometrical terms, the one qubit
transformation can be thought of as rotating the coordinate system of the
initial Bloch sphere of qubit $a$ before the interaction takes place, thus
changing which initial pure states are mapped to particular points on the
ellipsoids for the outgoing qubits after the interaction.  The transformation
$S_a$ performs some rotation (real orthogonal transformation with determinant
$+1$) on the Bloch sphere for the outgoing qubit $a$, thus rotating the
corresponding ellipsoid as a rigid body, leaving its semi-axes, as well as its
distance from the center of the Bloch sphere unchanged.  Of course $S_b$
performs a similar operation on the Bloch sphere of the outgoing qubit $b$.

		\section{Stochastic copying machine}
\label{sx}

	Are there cases in which two copies of a single qubit can be produced
by a copying machine using ancillary qubits (in addition to the two qubits
required for the copies), but the same task cannot be carried out by the
two-qubit copier described above?  In \cite{r2} we showed how to construct an
optimal copying machine using three qubits (one ancillary qubit) in which the
output ellipsoids for the two copies are centered in their respective Bloch
spheres, leading to symmetrical copying errors. This is clearly not possible,
at least in general, for a two-qubit copier, since the fact that the output
ellipsoids are tangent to the corresponding Bloch spheres, see Fig.~\ref{f1},
means that they will be off center.  However, as we shall show, it is possible
to move these ellipsoids to the centers of their respective Bloch spheres by
means of a {\it stochastic} two-qubit copying machine \cite{r7}.  While such a stochastic
machine can, obviously, produce different results (in a statistical sense) from
a simple copier, we do not know its limitations. 

	A stochastic copying machine is one for which certain parameters
entering the unitary transformation can be varied in a random way.  For
example, imagine two two-qubit copying machines of different construction, with
the input sometimes fed to one machine and sometimes to the other, the choice
being made at random. Alternatively, a single machine may be equipped with a
switch which can be randomly flipped between one of two different positions,
producing two different unitary transformations inside the machine which result
in different isometries.  In either case, one can suppose that the choice
between the two possibilities is generated by some ``classical'' random number
generator, although a ``quantum coin'' could also be employed, as we shall see.
Obviously, one could imagine using three or more machines, or one machine with
a switch which could be set at three or more positions, but for simplicity we
shall restrict our discussion to the case where there are only two
possibilities.

	For a given $\rho^{in}$, suppose the copy machine produces an output
$\rho^f_0$ in channel $f=a$ or $b$ for a switch setting of 0, and $\rho^f_1$ for
a switch setting of 1, corresponding in Bloch sphere language, see
(\ref{e3.4}), to
\beq
\rbd^f_i=\Mbd^f_i\cdot \rbd^{in}+\dbd^f_i,
\label{ex.1}
\eeq
with $i=0,1$.  Given that setting 0 occurs with probability $p_0$, and 1 with
probability $p_1=1-p_0$, the density matrices for the output channels will be
given by
\beq
\bar\rho^f=p_0\rho^f_0+p_1\rho^f_1,
\label{ex.2}
\eeq
corresponding to
\beq
\bold{\bar r}^f=\bold{\bar M}^f\cdot \rbd^{in}+\bold{\bar d}^f,
\label{ex.3}
\eeq
with
\beq
 \bold{\bar M}^f = p_0 \Mbd^f_0 + p_1 \Mbd^f_1,\quad \bold{\bar d}^f = p_0
\dbd^f_0 + p_1 \dbd^f_1.
\label{ex.4}
\eeq
Of course, precisely the same formulas apply if one imagines two distinct
copying machines rather than a single copying machine with a switch.

	As a specific example, let us suppose that the switch settings $i=0$
and 1 result in a canonical isometry specified by (\ref{e3.5}) to (\ref{e3.7}),
with $\dl=\dl_i$ and $\gm=\gm_i$ for setting $i$, and with the two pairs of
angles related by
\beq
 \gm_1=\gm_0,\quad \dl_1 = \pi-\dl_0.
\label{ex.5}
\eeq
In addition, for $i=1$, but not for $i=0$, a unitary transformation $|0\rg \rar
|1\rg$, $|1\rg\rar -|0\rg$, equivalent to $R(\pi)$ in (\ref{e4.2}), is applied
to the $b$ output, i.e., as $S_b$ in (\ref{e2.10}).  Since this amounts to a
rotation of the Bloch sphere by $\pi$ about the $y$ axis ($x\rar -x$,
$z\rar -z$), one can easily convince oneself that
\beq
 \Mbd^f_1 = \Mbd^f_0,\quad \dbd^f_1 = -\dbd^f_0.
\label{ex.6}
\eeq
Consequently, (\ref{ex.4}) tells us that the output of this stochastic copying
machine is the same as that of the non-stochastic machine with $i=0$, except
that the ellipsoid for each channel is displaced along the $z$ axis in the
corresponding Bloch sphere by an amount which depends upon $p_0$.  In
particular, when $p_0=p_1=1/2$, the corresponding ellipsoids are centered.

	This result indicates that at least in certain circumstances a
stochastic two-qubit copier can do the same job as a (more expensive) copier
using additional, ancillary qubits.  However, it is not clear that this is
always the case.  The construction just employed produces centered ellipsoids
representing an optimal copying machine according to the criteria of
\cite{r2}, but these ellipsoids have a special relationship among their
principal axes which does not have to hold for a more general copier.  On the
other hand, the copier described above represents only one among a very large
number of possible stochastic machines.  What stochastic copiers can and cannot
accomplish is a problem which remains to be explored.

	It is easy to show that a stochastic machine of the form we are
considering can be replaced by an equivalent non-stochastic or ``unitary''
copier employing an ancillary qubit.  We suppose that the ancillary
qubit $c$ is initially in the state
\beq
\sqrt{p_0}|0_c\rg+\sqrt{p_1}|1_c\rg,
\label{ex.7}
\eeq
and that the unitary time evolution for the copier  results in
\beqa
|\al\rg|0\rg|0_c\rg&\rightarrow&\left(U_0|\al\rg|0\rg\right)\otimes|0_c\rg,
\nn\\
|\al\rg|0\rg|1_c\rg&\rightarrow&\left(U_1|\al\rg|0\rg\right)\otimes|1_c\rg,
\label{ex.8}
\eeqa
where $U_0$ and $U_1$ are the two unitary transformations of the two-qubit
stochastic copier when $i=0$ and 1.  One can think of $|0_c\rg$ and $|1_c\rg$
as the two states of a ``quantum coin'', and using consistent history methods
of the type discussed in Sec.~\ref{s6c} below it is possible to produce a
quantum description of the copying process (\ref{ex.8}) in which the ancillary
bit is initially in $|0_c\rg$ with probability $p_0$ and $|1_c\rg$ with
probability $p_1$, so that the quantum coin controls the copying process in
very much the same was as a classical random number generator. (Alternatively,
one could prepare the quantum coin in the state (\ref{ex.7}), carry out a
measurement to determine whether it is in $|0_c\rg$ or $|1_c\rg$, and use the
resulting ``classical'' (macroscopic) signal to control a switch setting on the
copier.) 

	By generalizing this construction, using two or more ancillary qubits
in the case in which the switch has more than two positions, it is pretty
obvious that any stochastic copying machine can be replaced by a unitary
machine employing ancillary qubits.  Since stochastic machines with fewer
qubits are likely to be easier to construct than unitary machines employing
additional ancillary qubits, one would like to know under what conditions a
unitary copier can be replaced by a stochastic copier.  At present we do not
know the answer to this question, even in the simple case of a machine
producing two copies of one qubit.

		\section{Quantum circuit for isometry}
\label{s4}

	In the study of quantum computation, unitary transformations are often
written as a series of sub-transformations corresponding to simple operations
represented by quantum gates.  Doing this helps one better understand the
overall transformation, and suggests ways in which it might be implemented in
practice. It is known \cite{r8} that any unitary transformation on a
collection of qubits can be carried out using one-qubit gates, that is, unitary
transformations on a single qubit, together with a particular type of two-qubit
gate known as controlled-NOT or XOR.  One qubit gates should be much easier to
manufacture than two qubit gates, and thus there is an advantage to using as
few of the latter as possible.

\begin{figure}
\epsfxsize=10truecm
\hspace*{3cm}\epsfbox{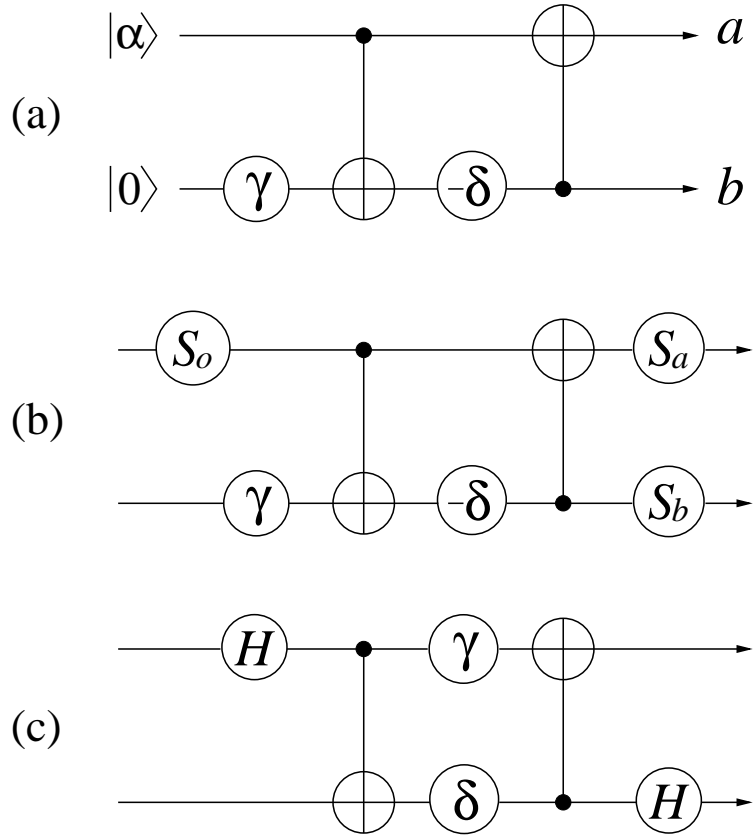}\vspace{.5cm}
\caption{
Quantum circuits: (a) The canonical form $V_c$, (\protect\ref{e2.9}). (b) The
isometry $V$ in (\protect\ref{e2.10}). (c) Alternative circuit producing the
same isometry as (a).  The gates are defined in (\protect\ref{e4.1}) to
(\protect\ref{e4.3}).}
\label{f2}
\end{figure}

	The canonical isometry (\ref{e2.9}) can be implemented as shown in
Fig.~\ref{f2}(a) using two controlled-NOT gates together with two one-qubit
gates.  The upper and lower horizontal lines in the figure correspond to the
$a$ and $b$ qubits, thought of as moving in parallel through the circuit from
left to right as time progresses.  Initially the $b$ qubit is in the state
$|0\rg$, whereas the $a$ qubit is in an arbitrary state $|\al\rg$, and the
circuit produces the transformation (\ref{e2.9}) on the initial $a$ qubit.  The
controlled-NOT gate is denoted by a vertical line between a {\it controlling}
qubit, indicated by a solid dot, and the {\it controlled} qubit, indicated by a
circled cross.  In the first (left) controlled-NOT gate in the figure, the $a$
qubit is the controlling qubit, and the unitary transformation represented by
the gate on the pair $|ab\rg$ is defined by
\beqa
|00\rg\rar|00\rg,&\quad&|10\rg\rar |11\rg,\nn\\
|01\rg\rar|01\rg,&\quad&|11\rg\rar |10\rg,
\label{e4.1}
\eeqa
using the same notation as in (\ref{e2.9}).  In the second controlled-NOT gate
in Fig.~\ref{f2}(a), $b$ is the controlling qubit, so (\ref{e4.1}) applies with
the arguments of each ket interchanged.

	The open circles in Fig.~\ref{f2}(a) represent one-qubit gates
producing a unitary transformation of the form
\beqa
R(\phi)|0\rg&=&+\cos(\phi/2) |0\rg+\sin(\phi/2) |1\rg,\nonumber\\
R(\phi)|1\rg&=&-\sin(\phi/2) |0\rg+\cos(\phi/2) |1\rg.
\label{e4.2}
\eeqa
In Bloch sphere language this is a rotation by an angle $\phi$ about the $y$
axis. The value of $\phi$ associated with each transformation in Fig.~\ref{f2}
is indicated inside the circle, and the angles $\gm$ and $\dl$ are related
through (\ref{e3.8}) to $\zt$ and $\eta$ in (\ref{e2.9}).  The isometry $V$
whose canonical form is $V_c$ can then be produced by adding to the circuit in
Fig.~\ref{f2}(a) the additional one-qubit gates shown in (b), corresponding to
the unitary transformations in (\ref{e2.10}).

	It is important to note that many different unitary transformations,
and thus many different quantum circuits, can produce the same isometry. (In
addition, the same unitary transformation can be produced by more than one
circuit.)  For example, the circuit in Fig.~\ref{f2}(c) represents a different
unitary transformation from the one in (a), but produces precisely the same
isometry on the initial $a$ qubit.  The one-qubit Hadamard gate labeled $H$
corresponds to a transformation
\beqa
 H|0\rg&=& (|0\rg+ |1\rg)/\sqrt{2},
\nonumber\\
 H|1\rg&=& (|0\rg- |1\rg)/\sqrt{2}.
\label{e4.3}
\eeqa
As these Hadamard gates can be combined with the $S_0$ and $S_b$ gates when
producing the isometry $V$, Fig.~\ref{f2}(b), the circuit in (c) is effectively
no more complicated than the one in (a).

	There are, of course, many other circuits which can produce the same
isometry. In general such a circuit requires at least two controlled-NOT gates,
for if there is only one controlled-NOT present, such as the circuit in
Fig.~\ref{f2}(a) with the second two-qubit gate removed, the canonical form
$V_c$ of the isometry, (\ref{e2.9}), is restricted to the case $\zt=\eta$, or
something equivalent to this under the symmetries discussed in App.~\ref{sab}.
In Sec.~\ref{s6} we shall consider an example in which the possibility of
representing the same isometry by several different circuits can be a useful
source of physical insight.

	The stochastic copying machine described in Sec.~\ref{sx}, with the two
possibilities represented by (\ref{ex.5}) together with an added $R(\pi)$ on
the $b$ output when $i=1$, can be implemented by using the circuit of
Fig.~\ref{f2}(b), with $S_0$ and $S_a$ removed (that is, equal to the identity),
if the unitary transformations performed by two of the one-qubit gates can be
altered by means of a ``classical'' stochastic signal. For one value of the
signal, $S_b$ is the identity, while for the other value, $-\dl$ is changed to
$\dl-\pi$, and $S_b$ becomes $R(\pi)$.  If the signals come from a ``classical''
random number generator, this circuit will produce copies with errors of the
type described in Sec.~\ref{sx}.

	\section{Application to quantum eavesdropping}
\label{s6}

		\subsection{Introduction}
\label{s6a}

	In typical quantum cryptographic schemes \cite{r9} a quantum channel is
used to provide secure communication between two users, Alice and Bob, through
the fact that attempts by an eavesdropper Eve to obtain information by
inserting a copying machine into the channel will produce a detectable noise,
that is, a certain number of errors in the transmission from Alice to Bob.  The
problem of optimal eavesdropping is to determine how much information the
eavesdropper can obtain for a given maximum level of noise.

	Our purpose here is not to discuss the optimal eavesdropping problem.
For the cryptographic schemes we will be concerned with, the problem has already
been solved \cite{r5,r6} under the assumption that the eavesdropper makes 
a separate measurement for each of the signals sent from Alice to Bob.  
Instead, we will show that the optimal eavesdropping schemes which
have been proposed can be carried out using simpler, thus ``cheaper'', quantum
circuits than were previously known.  In particular, the two qubit copying
machine of the present paper can be used in place of the three qubit machine
discussed previously in \cite{r10} for the BB84 \cite{r3} scheme, and a
certain simplification of the two-qubit eavesdropping machine in \cite{r6}
suffices for the B92 \cite{r4} scheme.

		\subsection{The BB84 cryptographic scheme}\
\label{s6b}

	In the BB84 protocol, Alice sends Bob one qubit at a time through a
quantum channel, using one of two basis vectors, chosen randomly, belonging to
one of two orthonormal bases or {\it modes}, and the mode is also chosen at
random.  It will be convenient, in view of the notation employed earlier in
this paper, to assume that these modes are an $x$ mode with basis vectors
\beq
|0_x\rg=\left(|0\rg+|1\rg\right)/\sqrt{2},\quad
|1_x\rg=\left(|0\rg-|1\rg\right)/\sqrt{2},\label{e6.1}
\eeq 
and a $y$ mode with basis vectors
\beq
|0_y\rg=\left(|0\rg+i|1\rg\right)/\sqrt{2},\quad
|1_y\rg=\left(|0\rg-i|1\rg\right)/\sqrt{2}.\label{e6.2}
\eeq
Eve, who does not know which mode is being employed for any particular signal,
sets up a copying machine whose action is always the same, whatever the initial
qubit sent by Alice, and stores her copies for later measurement when Alice has
publicly announced the mode used for each transmission.

	It was shown in \cite{r5} that there is an inequality giving an upper
bound for the amount of information which Eve can gain about a signal sent in
one mode for a given amount of noise produced in the {\it other} mode, and in
\cite{r10} that the bound can be achieved using a copying circuit involving a
total of three qubits (the one sent by Alice, plus two additional ancillary
qubits provided by the copying machine).  For present purposes, it is simplest
to view the copying machine together with its input and output as constituting
a quantum channel from Alice to Bob and, at the same time, a quantum channel
from Alice to Eve, with the remaining qubits in each case thought of as
ancillary qubits.

	As long as both of these channels are {\it symmetrical} for the modes
of interest, which is to say the error rate is the same for sending $|0\rg$ or
$|1\rg$, the information-theoretic bound in \cite{r5} is equivalent to the
statement that if the noise or error rate for the $x$ mode in the Alice to Eve
channel is below a certain amount, then that for the $y$ mode in the Alice to
Bob channel must be above a certain amount.  That is, if Eve is learning a lot
about signals in the $x$ mode, the copying machine will produce a lot of noise
in $y$ signals sent from Alice to Bob.  In the Bloch sphere representation, the
error rate for a channel is given by the expression
\beq
D=1-\lg\al|\rho|\al\rg=(1-\rbd^{in}\cdot\rbd^{out})/2,
\label{e6.3}
\eeq
understood in the following way.  If an input signal $|\al\rg$, $\rbd^{in}$ in
the Bloch sphere representation, emerges from the channel described by a
density matrix $\rho$ corresponding to $\rbd^{out}$, and is measured in the
$+\rbd^{in},-\rbd^{in}$ basis, $D$ is the probability that this measurement
yields $-\rbd^{in}$, an error, rather than $+\rbd^{in}$.

	If the copying machine is described by the canonical isometry $V_c$,
the error rate $p_x$ for the Alice to Bob (output $a$) channel, and $q_x$ for
the Alice to Eve (output $b$) channel, for mode $x$, (\ref{e6.1}), are given by
\beq
 p_x=(1-\sin\gm)/2,\quad q_x=(1-\cos\dl)/2,
\label{e6.4}
\eeq
independent of whether a $|0_x\rg$ or a $|1_x\rg$ is transmitted, and for mode
$y$ by
\beq
 p_y=(1-\sin\dl)/2,\quad q_y=(1-\cos\gm)/2,
\label{e6.5}
\eeq
with the same error rate for $|0_y\rg$ and $|1_y\rg$.  The symmetry of the
error rates is a consequence of (\ref{e6.3}) and the fact that $\rbd^{in}$ lies
in the $x,y$ place, so it is only the projection of $\rbd^{out}$ in this plane
that matters.  Whereas the ellipsoids in Fig.~\ref{f1} are not centered in the
Bloch sphere, their projections on the $x,y$ plane are centered, and this
results in symmetrical error rates.

	From (\ref{e6.4}) and (\ref{e6.5}) it follows that
\beq
 (\half-q_x)^2 + (\half-p_y)^2 = 1/4,
\label{e6.6}
\eeq
which means that making $q_x$ small necessarily leads to $p_y$ large,
approaching $1/2$.  In geometrical terms, enlarging the $x$ semi-axis of the
$b$ qubit ellipsoid, Fig.~\ref{f1}, to improve the quality of eavesdropping in
mode $x$ necessarily results in a smaller $y$ semi-axis for the $a$ qubit, and
thus more noise for mode $y$ in the Alice-to-Bob channel. Of course,
\beq
 (\half-q_y)^2 + (\half-p_x)^2 = 1/4,
\label{e6.7}
\eeq
leads to the same sort of complementarity with the two modes interchanged.

	Precisely the same geometrical picture applies to the three-qubit
copying machine proposed in \cite{r2}, where the Bloch ellipsoids have the
same sizes and shapes as those in Fig.~\ref{f1}, for corresponding choices of
parameters, but are centered at the origin.  It is because the $x$
and $y$ modes are employed in the cryptographic protocol, and the corresponding
error rates remain unmodified when the ellipsoids are displaced from the origin
along the $z$ axis, that the two-qubit copying machine discussed in this paper
yields identical results to that of the three-qubit copier employed earlier.
Were signals also sent in the $z$ mode, corresponding to $| 0\rg$ and $| 1\rg$
in the notation of (\ref{e6.1}) and (\ref{e6.2}), the outputs of the two-qubit
and three-qubit copying machines would not be the same: in particular, the
errors would not be symmetrical for the $z$ mode.  Thus Alice and Bob might
want to make use of the $z$ mode as well as the $x$ and $y$ modes for detecting
eavesdropping.  However, Eve could employ a stochastic two-qubit copier,
as discussed in Sec.~\ref{sx}, to center the ellipsoids, thus symmetrizing the
$z$ mode error, so that there would still be no reason for her to invest in a
(presumably more expensive) three-qubit machine.

 		\subsection{Semiclassical analysis of errors for BB84}\
\label{s6c}

	In our previous discussion \cite{r10} of eavesdropping using a
three-qubit copier, we presented a simple ``semi-classical'' perspective for
understanding why increasing Eve's information gain for one mode necessarily
increases Bob's noise for the other mode.  A similar simple argument does not
exist for two-qubit copiers (or at least we have not found one), but a somewhat
more complicated version is possible if one exploits the possibility, mentioned
earlier, that several different unitary transformations, and thus different
quantum circuits, can produce the same isometry.

\begin{figure}
\epsfxsize=6truecm
\hspace*{5cm}\epsfbox{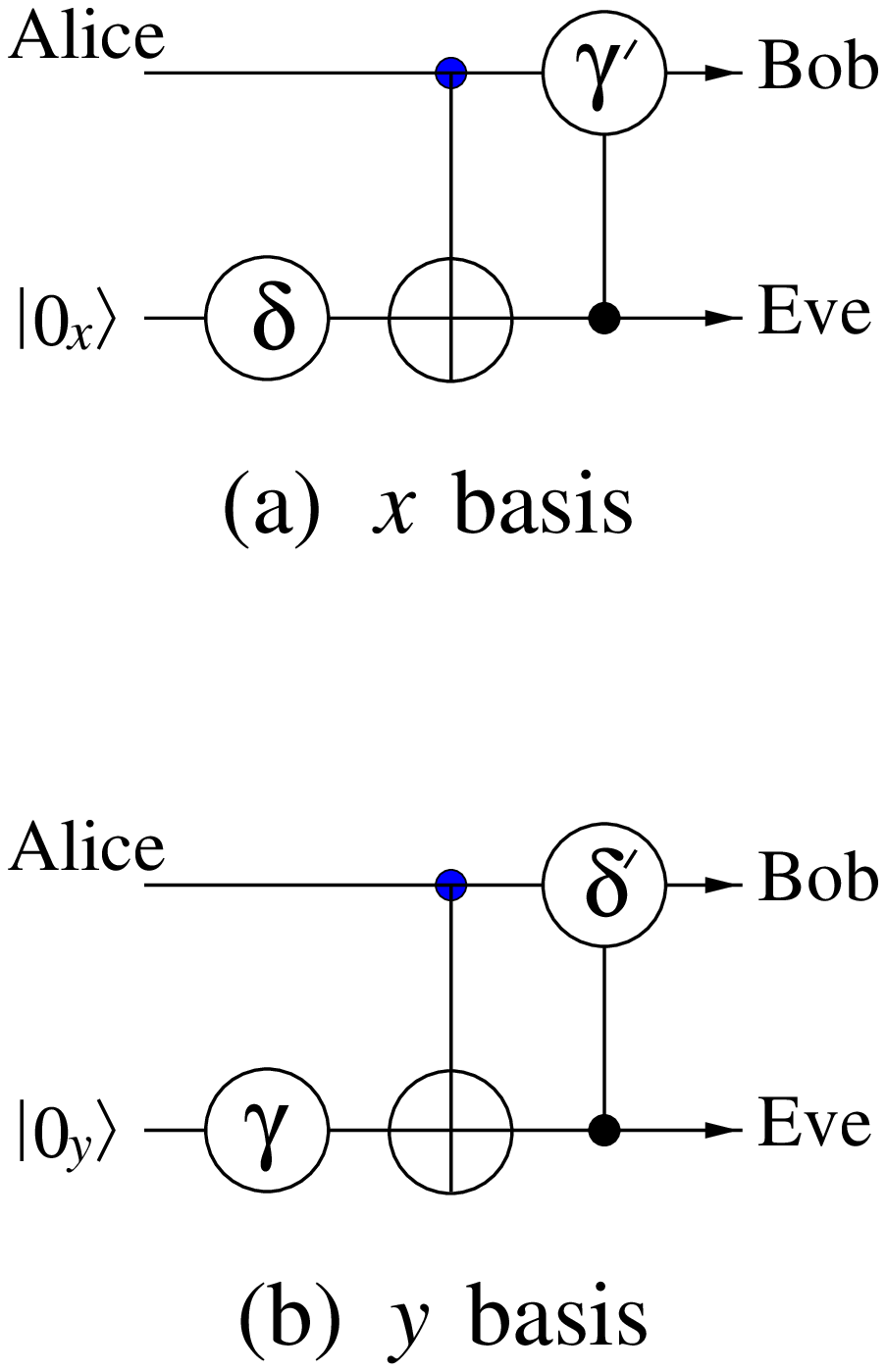}\vspace{.5cm}
\caption{
Two circuits which produce the same isometry as Fig.~\protect\ref{f2}(a).  The
circuit in (a) is in the $x$ basis, that in (b) is in the $y$ basis; see text.}
\label{f3}
\end{figure}

	In particular, the two circuits shown in Fig.~\ref{f3} produce the same
isometry as those in Fig.~\ref{f2}(a) or (c), when the angles $\gm'$ and $\dl'$
are defined to be
\beq
 \gm' = \pi/2 - \gm,\quad \dl' = \pi/2 - \dl,
\label{e6.8}
\eeq
and the action of the various gates is understood in the the following way.
	 In Fig.~\ref{f3}(a) the gates are to be thought of as in the $x$
representation using the basis (\ref{e6.1}).  In particular, the controlled-NOT
represents the transformation (\ref{e4.1}), but with $0$ and $1$ replaced
everywhere by $0_x$ and $1_x$, respectively.  Likewise, the action of the
one-qubit gate represented by $\dl$ inside a circuit is the same as
(\ref{e4.2}) provided $0$ and $1$ are replaced everywhere in (\ref{e4.2})---on
both sides---with $0_x$ and $1_x$, and $\phi$ is set equal to $\dl$.  Next, the
two-qubit gate following the controlled-NOT is a ``controlled-rotation''
producing the transformation
\beqa
|\al\rg|0_x\rg&\rar&\left[R(+\gm')|\al\rg\right]\otimes|0_x\rg,\nn\\
|\al\rg|1_x\rg&\rar&\left[R(-\gm')|\al\rg\right]\otimes|1_x\rg,
\label{e6.9}
\eeqa
where the one qubit gate $R(\phi)$ is (\ref{e4.2}) in the $x$ representation:
$0$ and $1$ replaced by $0_x$ and $1_x$.  Note that the ``rotation'' on qubit
$a$ is carried out in opposite senses depending upon whether $b$ is $|0_x\rg$
or $|1_x\rg$. Also, note that the initial state of the $b$ qubit is $|0_x\rg$,
rather than $|0\rg$, as in Fig.~\ref{f2}.

	The circuit in Fig.~\ref{f3}(b) is in the $y$ representation, which
means that it is to be interpreted in the same way as (a), but with $0_y$ and
$1_y$ substituted for $0_x$ and $1_x$.  Note that the circuits in Fig.~\ref{f3}
produce an isometry equivalent to Fig.~\ref{f2} with the definitions for the
$x$ and $y$ bases given in (\ref{e6.1}) and (\ref{e6.2}), but will not (at
least in general) do so for other basis choices using different phases.  For
example, using $|0_x\rg$ as defined in (\ref{e6.1}), but replacing $|1_x\rg$
with $(|1\rg - |0\rg)/\sqrt{2}$, and then using the circuit in Fig.~\ref{f3}(a)
will produce a very different result (try it!).

\begin{figure}
\epsfxsize=8truecm
\hspace*{4cm}\epsfbox{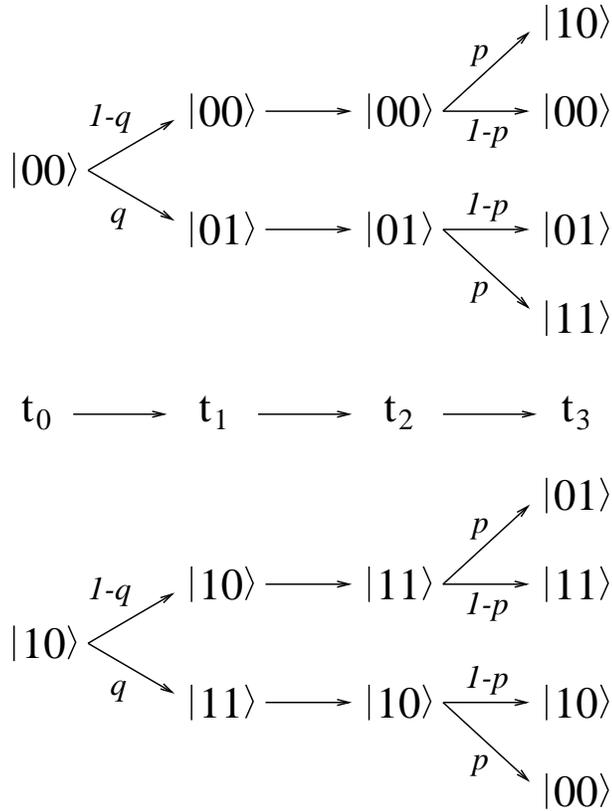}\vspace{.5cm}
\caption{
Family of eight quantum histories for the circuit in Fig.~\protect\ref{f3}(a).}
\label{f4}
\end{figure}

	Suppose Alice sends a signal using the $x$ mode, and Bob and Eve
eventually measure their qubits in the same mode.  Whatever two-qubit machine
Eve employs (e.g., that of Fig.~\ref{f2}(a)), we can, for purposes of obtaining
an intuitive picture of what is going on, imagine that copies are produced by
the circuit shown in Fig.~\ref{f3}(a), as this yields the same isometry.  This
circuit can be analyzed in the same manner as that in
\cite{r10}, using an appropriate family of quantum histories in which every
qubit is in either the $|0_x\rg$ or the $|1_x\rg$ state at times when it is not
inside one of the gates.  The different possible histories of this type are
indicated schematically in Fig.~\ref{f4}, where the $x$ subscripts have been
omitted. At time $t_0$ the two qubits have not passed through any gates, and
the two possible initial states, written as $|ab\rg$, are $|0_x0_x\rg$ and
$|1_x0_x\rg$. At $t_1$ the $b$ qubit has passed through the $\dl$ gate, so it
can be either $|0_x\rg$ or $|1_x\rg$; the probability of the latter is
\beq
 q_x = \left[\sin(\dl/2)\right]^2 = \half (1-\cos\dl).
\label{e6.10}
\eeq
This probability can be computed in the standard way using weights \cite{r11};
note that it is the same as what one would calculate if a measurement on this
qubit were to take place at time $t_1$ \cite{r12}.  By time $t_2$ the two qubits
have passed thought the controlled-NOT gate, while at $t_3$ the $a$ qubit has
passed through the $\gm'$ gate.  In passing through this gate, the $a$ qubit is
flipped, from $|0_x\rg$ to $|1_x\rg$, or vice versa, with a probability
\beq
 p_x = \left[\sin(\gm'/2)\right]^2 = \half (1-\sin\gm),
\label{e6.11}
\eeq
independent of whether the $b$ qubit is in the state $|0_x\rg$ or $|1_x\rg$,
since the probability only depends on the magnitude, not the sign of $\gm'$.

	The eight histories in Fig.~\ref{f4} constitute a family of quantum
histories which can be treated in the same way as a stochastic family of
classical histories, as long as quantum consistency conditions are satisfied
\cite{r11}. That these conditions are, indeed, satisfied for the family under
consideration follows from the fact that the two initial states are mutually
orthogonal, and for each initial state, all four final states at the right side
of Fig.~\ref{f4} are mutually orthogonal.  Hence the corresponding chain or
weight operators are orthogonal.

\begin{figure}
\epsfxsize=6truecm
\hspace*{5cm}\epsfbox{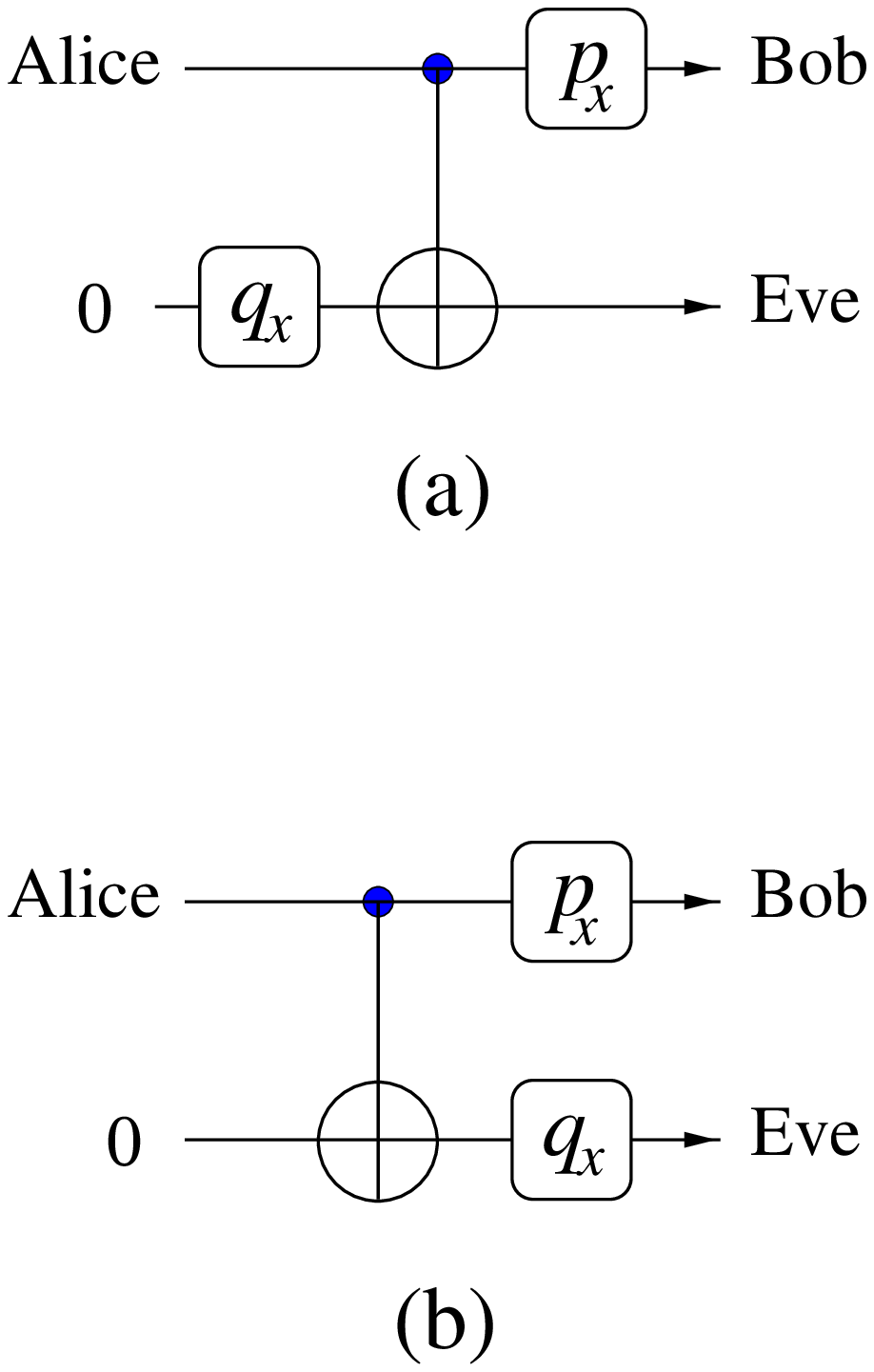}\vspace{.5cm}
\caption{
Classical stochastic circuits which are equivalent to Fig.~\protect\ref{f3}(a)
when using the consistent family of Fig.~\protect\ref{f4}.}
\label{f5}
\end{figure}
 
	Consequently, the action of the circuit in Fig.~\ref{f3}(a) is
precisely the same as that of a classical stochastic circuit shown in
Fig.~\ref{f5}(a), in which the gates labeled $p_x$ and $q_x$ correspond to
randomly flipping a bit from 0 to 1, or vice versa, with probabilities $p_x$
and $q_x$.  The effect of the $q_x$ gate is the same if it is placed after,
rather than before, the controlled-NOT, as in Fig.~\ref{f5}(b), which makes the
action of the circuit perfectly transparent: Since the $b$ bit is initially 0,
the controlled-NOT copies the $a$ bit, 0 or 1, to the $b$ bit.  Then both bits
are randomly flipped with appropriate probabilities before the $a$ bit is
measured by Bob and the $b$ bit by Eve.

	It is evident that the eavesdropper will obtain the most information
possible when $q_x=0$, while $p_x=0$ yields the minimum amount of noise in the
Alice-to-Bob channel.  Thus, choosing $\dl=0$ and $\gm=\pi/2$ represents the
optimal eavesdropping strategy if the $x$ mode is considered by itself.  But
these values create problems when Alice transmits using the $y$ mode.  To
analyze what happens in this case, it is convenient to imagine that the actual
copying machine, whatever it may be, is replaced by the circuit in
Fig.~\ref{f3}(b), which produces the same isometry.  Our preceding analysis of
part (a) of that figure can be applied to (b) by simply replacing $x$ with $y$
and noting the difference in the gate parameters in the two cases.  Thus the
consistent family is that of Fig.~\ref{f4}, where one now understands the
symbols as having $y$ subscripts, and the action of the quantum circuit, for the
$y$ mode, is the same as that of the classical circuits in Fig.~\ref{f5}, with
$p_x$ and $q_x$ replaced by
\beqa
  p_y& =& \left[\sin(\dl'/2)\right]^2 = \half (1-\sin\dl),
\label{e6.12}\\
 q_y & =& \left[\sin(\gm/2)\right]^2 = \half (1-\cos\gm).
\label{e}
\eeqa
Consequently, the choice $\dl=0$, which provides Eve with the optimal amount of
information about signals in the $x$ mode, creates the maximum possible amount
of noise in the Alice-to-Bob channel when used in the $y$ mode.  Likewise, if
Eve chooses $\gm=\pi/2$ in order to remain ``invisible'', producing no noise,
when Alice is transmitting and Bob is measuring in the $x$ mode, the
consequence will be that she gains no information at all when Alice transmits
in the $y$ mode.

\begin{figure}
\epsfxsize=6truecm
\hspace*{5cm}\epsfbox{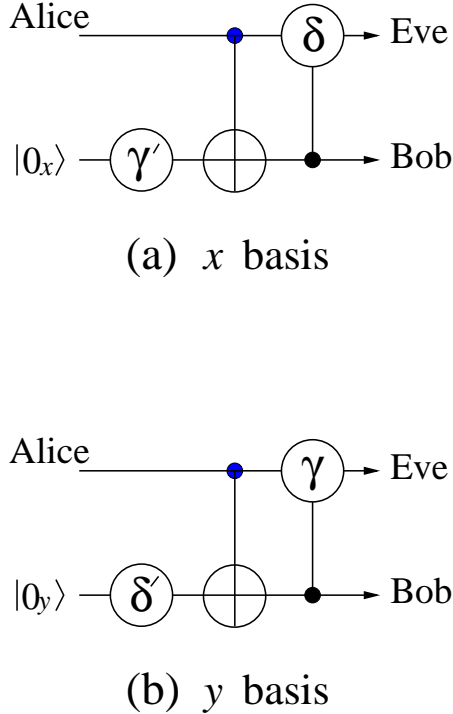}\vspace{.5cm}
\caption{
A circuit (a) in the $x$ basis and a different circuit (b) in the $y$ basis
which produce the same isometry as the circuits in Fig.~\protect\ref{f2}(a),
(c), and Fig.~\protect\ref{f3}.}
\label{f6}
\end{figure}

	Many other circuits will produce the same isometry on the $a$ qubit as
those in Fig.~\ref{f2}(a) and Fig.~\ref{f3}(a) and (b). For example, the two
shown in Fig.~\ref{f6}(a) and (b) are similar to those shown in Fig.~\ref{f3},
except that the destination of the final bits has been interchanged.  The four
circuits in Figs.~\ref{f3} and \ref{f6} represent {\it different} unitary
transformations, and thus an actual copying machine could employ only one of
these circuits.  Hence it is worth emphasizing, once again, that all of these
circuits, because they yield the same isometry, produce precisely the same
result in terms of copies of whatever Alice sends, in the $x$ or $y$ or in any
other mode.  However, some circuits are more useful than others when one wants
to form an intuitive picture of why certain parameter values lead to particular
sorts of errors.

		\subsection{The B92 cryptographic scheme}
\label{s6d}

	In the B92 cryptographic protocol, Alice sends signals through a
quantum channel in one of two non-orthogonal states, $|\kp_0\rg$ and
$|\kp_1\rg$, chosen at random, while Bob makes measurements in one of two
orthonormal bases, also chosen at random, one of which includes $|\kp_0\rg$ and
the other $|\kp_1\rg$. For details, see \cite{r6}.  Once again, eavesdropping is
detected through the production of noise that can be detected by the legitimate
users of the channel.

	Fuchs and Peres \cite{r6} carried out a numerical and analytic study of
the problem of distinguishing these two non-orthogonal states, and found an
optimal strategy (under certain plausible assumptions) for doing this with
minimal disturbance to the original signal, the one sent on to Bob.  This
strategy can be carried out using the two-qubit copying circuit in
Fig.~\ref{f3}(a).  Let the two non-orthogonal states used by Alice be
\beqa
|\kp_0\rg&=&\cos\bar\al|0_x\rg+\sin\bar\al|1_x\rg,\nn\\
|\kp_1\rg&=&\sin\bar\al|0_x\rg+\cos\bar\al|1_x\rg,
\label{e6.14}
\eeqa
with $\bar\al$ ($\al$ in \cite{r6}) a parameter which determines the degree of
non-orthogonality.  Eve carries out measurements in the $x$ basis
$|0_x\rg,|1_x\rg$, and uses this information to try and determine whether Alice
transmitted $|\kp_0\rg$ or $|\kp_1\rg$.

	The copying machine in Fig.~\ref{f3}(a) has two parameters $\dl$ and
$\gm'$ which are chosen by Eve in the following way.  The value $\dl=0$
provides Eve with copies of the states (\ref{e6.14}) which are optimal in the
sense that her measurements provide the best possible discrimination between
them; perfect discrimination is not possible, because $|\kp_0\rg$ and
$|\kp_1\rg$ are not orthogonal.  Choosing some other value of $\dl$ reduces the
amount of information that Eve can gain about the transmitted signals.  For a
given $\dl$, Eve chooses $\gm'$ in such a way as to minimize the noise produced
in the Alice-to-Bob circuit.  The appropriate value is worked out in \cite{r6},
where the parameters $\phi$ and $\theta$ are related to our $\dl$ and $\gm'$ by
\beq
 \phi=\dl/2,\quad \theta = \gm'/2.
\label{e6.15}
\eeq
While the optimal $\gm'$ is a somewhat complicated function of $\dl$, it turns
out that by increasing $\dl$ and thereby reducing the amount of information she
obtains, Eve can also reduce the amount of noise detectable by Alice and Bob.
Thus the optimal strategy can be thought of either as obtaining the most
information for a given level of noise, or producing the least amount of noise
(by a suitable choice of $\gm'$) for a given amount of information (determined
by the choice of $\dl$).

\begin{figure}
\epsfxsize=10truecm
\hspace*{3cm}\epsfbox{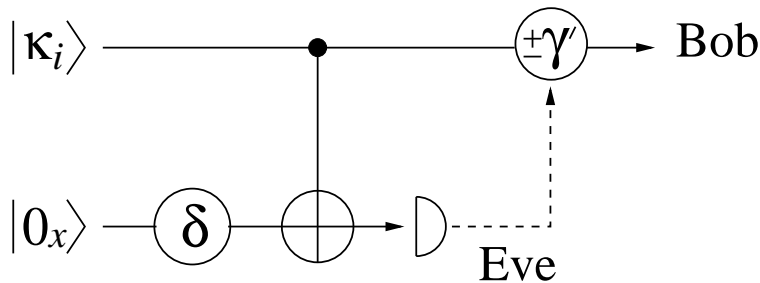}\vspace{.5cm}
\caption{
Improved eavesdropping machine for the B92 protocol, in which the last gate in
Fig.~\protect\ref{f3}(a) has been replaced by two single-qubit operations
determined by Eve's measurement result: $R(\gm')$ for $|0_x\rg$ and $R(-\gm')$
for $|1_x\rg$. Note that the gate notations and Eve's measurement are in the
$x$ basis, as in Fig.~\protect\ref{f3}(a).}
\label{f7}
\end{figure}

	The consistent family of Fig.~\ref{f4} is inappropriate for analyzing
this process because neither Alice nor Bob employ the $|0_x\rg,|1_x\rg$ basis.
Nevertheless, because Eve measures her qubit in this basis, there is a
consistent quantum-mechanical description in which Eve's qubit is in one of
these two states at all times after it leaves the controlled-NOT gate in
Fig.~\ref{f3}(a).  Consequently, her measurement reveals a pre-existing value,
and the measurement can equally well be carried out before the qubit reaches
the controlled-rotation gate, as in the modified circuit in Fig.~\ref{f7}.  In
this circuit the results of the measurement of $|0_x\rg$ or $|1_x\rg$ are used
to produce a ``classical'' signal which controls the operation of a one qubit
gate that carries out a unitary transformation, $R(+\gm')$ or $R(-\gm')$ as
appropriate, on the qubit sent on to Bob. This use of retrodiction for
simplifying a quantum circuit has been employed previously to simplify the
Fourier transform in Shor's factorization algorithm
\cite{r13}, and to study teleportation \cite{r14}.  Its justification lies in
a correct application of consistent-history methods \cite{r11}.

	Since one-qubit gates, even those controlled by classical signals, are
likely to be much easier to construct than two qubit gates, the circuit in
Fig.~\ref{f7} represents a cost-effective approach to eavesdropping in the case
of the B92 scheme.  Unfortunately (or perhaps fortunately), it cannot be used
for the BB84 protocol discussed earlier, because in that case Eve does not know
ahead of time which mode Alice will employ for transmitting signals, and
therefore has to wait until after the copying process is complete before
deciding whether to carry out a measurement in the $x$ or the $y$ basis.

		\section{Summary and open questions}
\label{s7}

	We have carried out a general analysis of a unitary transformation on
two qubits from the point of view of a copying machine producing two copies of
an arbitrary state of one of the input qubits while the other is held fixed.
Such a transformation produces an isometry mapping the input qubit onto a two
dimensional subspace $\Gc$ of the tensor product $\Ac\ot\Bc$ of the two qubits.
Central to our analysis is the theorem in Sec.~\ref{s2a} according to which
there is a basis for $\Gc$ in the canonical form given by (\ref{e2.5}).  In
some sense this is an extension, to a particular case, of the familiar Schmidt
representation for a vector on an arbitrary tensor product $\Ac\ot\Bc$.  Using
this canonical basis allows us to write an arbitrary isometry as a relatively
simple canonical isometry, depending on two real parameters, preceded and
followed by some additional unitary operations on individual qubits.

	A Bloch sphere representation of the output qubits, Sec.~\ref{s3},
provides a geometrical picture which is very useful for understanding what
sorts of copying errors can occur and the limitations on copy quality imposed
by quantum theory.  Despite its simplicity (or perhaps because of it), a
two-qubit copying machine produces optimal copies according to the criteria
worked out in \cite{r2}.  By using a stochastic generalization,
Sec.~\ref{sx}, one can obtain greater flexibility in determining the types of
errors which are produced by the machine.

	An isometry can be carried out using one of several possible quantum
circuits constructed from one-qubit gates and controlled-NOT two-qubit gates,
Sec.~\ref{s4}. As two-qubit gates are likely to prove rather expensive, it is
worth noting that two of them are sufficient, and also necessary for a general
isometry.  In addition, the canonical isometry corresponding to the canonical
basis choice for $\Gc$ requires two one-qubit gates, whereas a general isometry
requires a total of five one-qubit gates.  Some gates controlled by a
``classical'' stochastic signal are needed to implement a stochastic copying
machine. 

	The two-qubit copier suffices for carrying out the optimal
eavesdropping strategy derived in \cite{r5} for the BB84 cryptographic scheme.
The Bloch sphere picture is particularly valuable in showing why this simpler
copier can perform as well as the three-qubit machines proposed in
\cite{r2,r10}.  The stochastic two-qubit
copier will also do as well as a three-qubit copier for certain generalizations
of BB84.  Physical insight into how the parameters of the two-qubit machine
determine the information gained by the eavesdropper and the noise produced in
the channel between the legitimate users is obtained by employing consistent
families of quantum histories in two different circuits producing the same
isometry.

	A two-qubit copier which carries out the optimal eavesdropping strategy
for the B92 cryptographic protocol proposed by Fuchs and Peres
\cite{r6} can be further simplified by replacing the  second controlled-NOT gate
by a one-qubit gate controlled by the classical signal produced by an earlier
measurement.  This trick for producing a more economical machine relies upon
the justification of quantum retrodiction provided by a consistent history
analysis as first pointed out in \cite{r13}.

	While we have been able to characterize the most general isometry from
one qubit to a tensor product space $\Ac\ot\Bc$ of two qubits, this is not the
same thing as describing the most general unitary transformation on
$\Ac\ot\Bc$, where work remains to be done \cite{r15}.  Generalizing the
results of this paper to spaces $\Ac$ and $\Bc$ of dimension greater than two
represents a challenging problem.  One step towards solving it would be to find
some counterpart of the canonical basis, (\ref{e2.5}), when $\Gc$, $\Ac$, and
$\Bc$ are of any dimension.  Another would be the development of some
geometrical analogy of the Bloch sphere picture for higher-dimensional spaces.
And extensions to tensor products of three or more spaces of any of our results
would represent a significant contribution to quantum information theory.  The
possibilities and limitations of stochastic copying machines are worth
exploring, given that in certain circumstances they appear to offer a more
economical solution to copying problems than employing additional ancillary
qubits in a quantum circuit. 

\section*{Acknowledgments}

	One of us (CSN) thanks C. Fuchs for helpful conversations. Financial
support for this research has been provided by the NSF and ARPA through grant
CCR-9633102.

\appendix
		\section{The canonical representation}
\label{saa}

	Given a state $|\psi\rg$ of $\Ac\ot\Bc$ in the form
\beq
 |\psi\rg=\sum_{jk} \tau_{jk} |a'_jb'_k\rg,
\label{ea.1}
\eeq
where $\{|a'_0\rg,|a'_1\rg\}$ and $\{|b'_0\rg,|b'_1\rg\}$ are any orthonormal
bases of $\Ac$ and $\Bc$, it is easy to show that a necessary and sufficient
condition for $|\psi\rg$ to be a product state of the form $|a\rg\ot |b\rg$ is
that
\beq
 \tau_{00}\tau_{11} = \tau_{01}\tau_{10}.
\label{ea.2}
\eeq

	The lemma that a two-dimensional subspace $\Gc$ of $\Ac\ot\Bc$ always
contains a non-zero product state can be established as follows.  Assume that
$|\psi'\rg$ and $|\psi''\rg$ are linearly independent vectors in $\Gc$, and
$|\psi'\rg$ is not a product state. Applying the condition (\ref{ea.2}) to
\beq
 |\psi\rg = \lam |\psi'\rg + |\psi''\rg = 	\sum_{jk} (\lam\tau'_{jk} +
\tau''_{jk}) |a'_j b'_k\rg
\label{ea.3}
\eeq
leads to a quadratic equation in $\lam$ with a non-zero coefficient of
$\lam^2$, which thus has at least one root corresponding to a product state.
(In general one expects two roots and two linearly-independent product states.)
Consequently, there is an orthonormal basis
\beq
 |\bar g_0\rg = |\bar a_0 \bar b_0\rg,\quad |\bar g_1\rg = c_{01}|\bar a_0 \bar
b_1\rg + c_{10}|\bar a_1 \bar b_0\rg + c_{11}|\bar a_1 \bar b_1\rg
\label{ea.4}
\eeq
of $\Gc$ written in terms of suitable orthonormal bases $\{|\bar a_j\rg\}$ and
$\{|\bar b_j\rg\}$ of $\Ac$ and $\Bc$, where $|\bar g_0\rg$ is the product
state whose existence is guaranteed by the lemma.  If $c_{11}=0$, removing the
bars in (\ref{ea.4}) yields (\ref{e2.5}) with $\bar \mu=0$, and our task is
finished.

	Thus from now on we assume that
\beq
 c_{11} \neq 0.
\label{ea.5}
\eeq
There is then a second product state
\beq
 |\psi\rg = (c_{01}c_{10}/c_{11}) |\bar g_0\rg + |\bar g_1\rg,
\label{ea.6}
\eeq
in $\Gc$---as can be verified using (\ref{ea.2})---which is linearly
independent of $|\bar g_0\rg$. It is convenient to write these two
linearly-independent product states in the normalized (but not orthogonal) form
\beq
 |\pi_j\rg = |\al_j\rg\ot |\bt_j\rg,
\label{ea.7}
\eeq
for $j=0,1$, with
\beq
 \lg \al_j | \al_j \rg = 1 = \lg \bt_j | \bt_j \rg,
\label{ea.8}
\eeq
and with phases chosen so that
\beq
 0 \leq \lg \al_0 | \al_1 \rg < 1,\quad 0 \leq \lg \bt_0 | \bt_1 \rg < 1.
\label{ea.9}
\eeq
The inequalities in (\ref{ea.9}) are strict, since were it the case, for
example that, $\lg \al_0 | \al_1 \rg = 1$, this would mean
$|\al_0\rg=|\al_1\rg$, and $\Gc$ would consist entirely of product states of
the form $|\al_0\rg\ot |b\rg$, which is inconsistent with (\ref{ea.5}).

	Because of the strict inequality just noted, none of the four kets
\beqa
 && |\hat a_0\rg = |\al_0\rg+|\al_1\rg,\quad |\hat a_1\rg =
|\al_0\rg-|\al_1\rg,
\nn\\
 && |\hat b_0\rg = |\bt_0\rg+|\bt_1\rg,\quad |\hat b_1\rg = |\bt_0\rg-|\bt_1\rg
\label{ea.10}
\eeqa
is zero, and since each pair is orthogonal,
\beq
 \lg \hat a_0 | \hat a_1 \rg = 0 = \lg \hat b_0 | \hat b_1 \rg,
\label{ea.11}
\eeq
one can construct orthonormal bases for $\Ac$ and $\Bc$ by appropriate
normalization:
\beq
 |a_j\rg = |\hat a_j\rg / \sqrt{\lg a_j | a_j \rg },\quad |b_j\rg = |\hat
b_j\rg / \sqrt{\lg b_j | b_j \rg }.
\label{ea.12}
\eeq
Inverting the relations (\ref{ea.10}) allows one to write
\beqa
 && |\al_0\rg = \kp|a_0\rg + \bar \kp |a_1\rg,\quad |\al_1\rg = \kp|a_0\rg -
\bar \kp |a_1\rg,
\nn\\
 && |\bt_0\rg = \lam|b_0\rg + \bar \lam |b_1\rg,\quad |\bt_1\rg = \lam|b_0\rg -
\bar \lam |b_1\rg,
\label{ea.13}
\eeqa
where the constants $\kp,\bar\kp,\lam,\bar\lam$ are all strictly positive
numbers.  Consequently, the two orthogonal vectors
\beqa
 && |\hat g_0\rg = |\pi_0\rg + |\pi_1\rg = 2(\kp \lam|a_0b_0\rg + \bar\kp
\bar\lam|a_1b_1\rg),
\nn\\
  && |\hat g_1\rg = |\pi_0\rg - |\pi_1\rg = 2(\kp \bar\lam|a_0b_1\rg + \bar\kp
\lam|a_1b_0\rg),
\label{ea.14}
\eeqa
when appropriately normalized, provide an orthonormal basis for $\Gc$ in the
form (\ref{e2.5}).

	\section{Symmetries of coefficients in the canonical representation}
\label{sab}

	The representation (\ref{e2.5}) is in general not unique in that the
same subspace $\Gc$ may possess an alternative orthonormal basis
\beqa
 |g'_0\rg &=& \mu' |a'_0 b'_0\rg + \bar \mu' |a'_1 b'_1\rg,
\nn\\
 |g'_1\rg &=& \nu' |a'_0 b'_1\rg + \bar \nu' |a'_1 b'_0\rg,
\label{eb.1}
\eeqa
written using alternative orthonormal bases $\{|a'_0\rg,|a'_1\rg\}$ for $\Ac$
and $\{|b'_0\rg,|b'_1\rg\}$ for $\Bc$.  Different sets of coefficients
$\{\mu',\bar\mu',\nu',\bar\nu'\}$ which can be used to represent the same
subspace will be called {\it equivalent}, and maps which carry one set of
coefficients to an equivalent set will be referred to as {\it symmetry
operations}.  (Note that replacing the coefficients in (\ref{e2.5}) by an
equivalent set while leaving the bases for $\Ac$ and $\Bc$ unchanged will, in
general, lead to a different subspace $\Gc$; the coefficient change must be
accompanied by changes in the bases if $\Gc$ is to remain unaltered.)

	It is easy to show that multiplying the coefficients by arbitrary phase
factors,
\beq
 \mu'=e^{i\phi_{00}}\mu,\quad \bar\mu'=e^{i\phi_{11}}\bar\mu,\quad
\nu'=e^{i\phi_{01}}\nu,\quad \bar\nu'=e^{i\phi_{10}}\bar\nu,
\label{eb.2}
\eeq
for any choice of $\phi_{jk}$, is a symmetry operation in the sense just
defined: insert
\beq
 |a'_j\rg = e^{i\al_j}|a_j\rg,\quad |b'_j\rg = e^{i\bt_j}|b_j\rg,\quad |g'_j\rg
= e^{i\gm_j}|g_j\rg,
\label{eb.3}
\eeq
in (\ref{eb.1}) and choose the six phases $\al_0,\al_1,\bt_0,\bt_1,\gm_0,\gm_1$
so as to recover (\ref{e2.5}).  (This is an alternative demonstration that the
coefficients in (\ref{e2.5}) can always be chosen to be real and positive.)

	Additional symmetries arise because it does not matter which of the
special basis vectors in $\Gc$ is called $|g_0\rg$ and which $|g_1\rg$;
likewise, one can interchange $|a_0\rg$ with $|a_1\rg$, or $|b_0\rg$ with
$|b_1\rg$.  These interchanges give rise to three symmetry operations on the
coefficients in addition to those in (\ref{eb.2}): (i) interchange $\mu$ with
$\nu$, and $\bar\mu$ with $\bar\nu$; (ii) interchange $\mu$ with $\bar\mu$, and
$\nu$ with $\bar\nu$; (iii) interchange $\mu$ with $\bar\nu$, and $\nu$ with
$\bar\mu$.  Of course, (iii) is just the product of (i) and (ii). (Note that
interchanging $\mu$ with $\bar\mu$ while keeping $\nu$ and $\bar\nu$ fixed is
{\it not} a symmetry operation.)

	When applied to the trigonometric representation (\ref{e2.7}), these
symmetry operations allow one to (i) change the sign of $\zt$; (ii) change the
sign of $\eta$; (iii) increase both $\zt$ and $\eta$ by $\pi$; (iv) interchange
$\zt$ with $\eta$.  (Note that increasing $\zt$ by $\pi$ without changing
$\eta$ is {\it not} a symmetry operation, although either $\zt$ or $\eta$ can
be increased by $2\pi$ while the other remains fixed.)  Combinations of these
operations map any $(\zt,\eta)$ in the square
\beq
 0\leq\zt\leq\pi,\quad 0\leq\eta\leq\pi,
\label{eb.4}
\eeq
corresponding to positive coefficients in (\ref{e2.7}), onto the equivalent
points
\beq
 (\eta,\zt), \quad (\pi-\zt,\pi-\eta), \quad (\pi-\eta,\pi-\zt),
\label{eb.5}
\eeq
from which it follows that $(\zt,\eta)$ can always, if desired, be chosen to
lie in the region (\ref{e2.8}).

	\section{Finding the canonical basis}
\label{sac}

	The construction in App.~\ref{saa} which demonstrates the existence of
the canonical representation is not an easy way to find it.  The following is
an alternative approach which is simpler and works except for certain
degenerate cases.  In particular, it does not require that one find product
states in $\Gc$.

	Given two linearly independent vectors in $\Gc$, one can construct
(Gram-Schmidt) an orthonormal basis $\{|\bar g_0\rg,|\bar g_1\rg\}$, and from
it the (unique) projector
\beq
 G = |\bar g_0\rg \lg \bar g_0| +|\bar g_1\rg \lg \bar g_1| = | g_0\rg \lg g_0|
+| g_1\rg \lg g_1|
\label{ec.1}
\eeq
onto the subspace $\Gc$, along with its partial traces
\beqa
 G_A &=& \Tr_{\Bc}[G] = (\mu^2 + \nu^2) |a_0\rg\lg a_0| + (\bar\mu^2 +
\bar\nu^2) |a_1\rg\lg a_1|,
\nn\\
 G_B &=& \Tr_{\Ac}[G] = (\mu^2 + \bar\nu^2) |b_0\rg\lg b_0| + (\bar\mu^2 +
\nu^2) |b_1\rg\lg b_1|,
\label{ec.2}
\eeqa
which are operators on $\Ac$ and $\Bc$, respectively.  Here we are assuming,
for convenience, that the coefficients in (\ref{e2.5}) are real and positive.

	If the eigenvalues of $G_A$, in parentheses on the right side of
(\ref{ec.2}), are non-degenerate, the dyads $|a_0\rg\lg a_0|$ and $|a_1\rg\lg
a_1|$ are uniquely defined up to identifying which is which. We can, for
example, assume that $|a_0\rg\lg a_0|$ corresponds to the larger and
$|a_1\rg\lg a_1|$ to the smaller eigenvalue.  A similar comment applies when
the eigenvalues of $G_B$ in (\ref{ec.2}) are non-degenerate.  The eigenvalues
of $G_A$ and $G_B$, along with the normalization condition (\ref{e2.6}), serve
to determine the non-negative coefficients $\mu,\bar\mu,\nu,$ and $\bar\nu$.

	The dyads $|a_0\rg\lg a_0|$, etc., in (\ref{ec.2}) determine vectors
$|a'_0\rg$, $|a'_1\rg$, $|b'_0\rg$, and $|b'_1\rg$ which are identical to their
unprimed counterparts in (\ref{e2.5}) apart from the arbitrary phase factors in
(\ref{eb.3}).  These phases {\it cannot} be chosen arbitrarily, because the
{\it relative} phases of the summands on the right side of (\ref{e2.5}) is
significant, and the information in the partial traces (\ref{ec.2}) is not
enough to determine them, so an additional step is needed.

	If $|g\rg$ is any vector in $\Gc$ (and hence a linear combination of
$|g_0\rg$ and $|g_1\rg$) such that both $\lg g |a_0 b_0\rg$ and $\lg g |a_1
b_1\rg$ are nonzero, then the positivity of the coefficients in (\ref{e2.5})
implies that
\beq
 \Ph(\lg g |a_0 b_0\rg) = \Ph(\lg g |a_1 b_1\rg),
\label{ec.3}
\eeq
where the phase $\Ph(z)$ is $\phi$ when $z=|z|e^{i\phi}$.  Given such a
$|g\rg$, which could be $|\bar g_0\rg$ or $|\bar g_1\rg$, the phases of the
inner products $\lg g |a'_0 b'_0\rg$ and $\lg g |a'_1 b'_1\rg$, along with
(\ref{ec.3}), constrain the choices of $\al_j$ and $\bt_j$ in (\ref{eb.3}). The
requirement that
\beq
 \Ph(\lg g |a_0 b_1\rg) = \Ph(\lg g |a_1 b_0\rg),
\label{ec.4}
\eeq
again based on (\ref{e2.5}), applied to a non-vanishing pair of inner products
$\lg g |a'_0 b'_1\rg$ and $\lg g |a'_1 b'_0\rg$, (with the same or a different
$|g\rg$ from that in (\ref{ec.3})) yields a second constraint for the $\al_j$
and $\bt_j$.  When both constraints are satisfied, the remaining freedom in
choosing phases simply determines the overall phases of $|g_0\rg$ and
$|g_1\rg$.


\begin{thebibliography}{99}
\bibitem{r12}
	While imagining that a measurement takes place can be an aid to
carrying out a calculation of the desired probability, it should be emphasized
that the probability is valid in the absence of any measurement provided the
consistency conditions for the family of histories are fulfilled.
\end{thebibliography}
\end{document}